# Effect of air confinement on thermal contact resistance in nanoscale heat transfer

Dheeraj Pratap,[1,a)] Rakibul Islam,[1,b)] Patricia Al-Alam,[1] Jaona Randrianalisoa[1] and Nathalie Trannoy [1,c)]

[1] *Groupe de Recherche en Sciences pour l'Ingenieur (GRESPI) – EA 4694, University of Reims Champagne-Ardenne, 51687 Reims Cedex 2, France*

**Abstract**

We report herein the pressure dependent thermal contact resistance ($R_c$) between Wollaston wire thermal probe and samples which is evaluated within the framework of an analytical model. This work presents heat transfer between the Wollaston wire thermal probe and samples at nanoscale for different air pressure ranging from 1 Pa to $10^5$ Pa. We make use of a scanning thermal microscopy (SThM) for the thermal analysis of two samples, fused silica ($SiO_2$) and Titanium (Ti), with different thermal conductivities. The thermal probe's output voltage difference ($\Delta V$) between out-off contact output voltage ($V_{oc}$) and in-contact output voltage ($V_{ic}$) was recorded. The result shows that the heat transfer increases with the increasing air pressure and it is found higher for higher thermal conductive material. We also propose an analytical model based on the normalized output voltage difference to extract the probe-sample thermal contact resistance. The obtained $R_c$ values in the range of $\sim 1.8 \times 10^7$ K/W to $\sim 14.3 \times 10^7$ K/W validate the presented analytical model. Further analysis reveals that the thermal contact resistance between the probe and sample decreases with increasing air pressure. Such behavior is interpreted by the contribution of heat transfer through confined air on thermal contact resistance. In addition, the signature of $R_c$ is also evidenced in the context of thermal mismatch behavior which is studied by using acoustic mismatch model (AMM) and diffuse mismatch model (DMM).

---

[a)]Electronic mail: dheerajpratap.sulr@gmail.com

[b)]Electronic mail: rakibul.islam@univ-reims.fr

[c)]Electronic mail: nathalie.trannoy@univ-reims.fr

# 1. Introduction

Heat transfer at nanoscale has been immense interest in last few decades because of the development of the nanotechnology and miniaturization of electronic devices[1–3]. At such small scale the heat management becomes critical which affects the reliability and efficiency of the devices[3,4]. To overcome these heat problems a good knowledge of heat distribution and more particularly localized hot points in devices and their management is needed at micro/nano scale. Unlike the optical techniques which suffer the diffraction limit, scanning thermal microscopy (SThM) has become important to study heat transport at very small scale[5–7]. The SThM is a kind of scanning probe microscopy (SPM) in which as special kind of probes, called as thermal probes, are used to study the heat transport at micro/nano scale. The thermal probes give topographic and thermal information simultaneously. The SThM has very high spatial resolution and thermal sensitivity[8]. Analysis of the heat exchange between the thermal probe and sample is a very active field of research to improve the heat measurement technique[9].

However, at nanoscale the heat transfer through interface between two materials becomes important due to thermal contact area acts as different thermal transport medium. More specifically, like strong contribution of interface zone on electrical transport[10,11], heat transfer is also significantly affected by interfacial thermal resistance[12–14]. Thermal contact resistance is the consequence of the imperfect matching of boundaries (roughness and asperities) and eventual third body such as, water meniscus, oxide layer and moisture etc[15,16]. In general, heat transfer is restricted to highly localised probe/sample contacts because of the nano and/or micro interfacial structures inducing thermal contact resistance. There are intense research on the thermal contact resistance at nanoscale but these were mainly performed at fixed pressure[9,17,18]. The contact resistance between the interface of thin film and samples are also reported where a high value of pressure (~0-50 GPa) was used[18]. But, there is no analysis available of varying air pressure effect on the thermal contact resistance of Wollaston thermal probe-sample interaction. It is, therefore, not surprising that much intense theoretical and applied research are still being conducted in hope to better understand the synergistic role played by thermal contact resistance.

The present work discusses an analysis of effect of surrounding air pressure on the thermal contact resistance between the thermal probe and sample. The air pressure in the vacuum chamber was varied from vacuum (1 Pa) to atmospheric pressure ($10^5$ Pa) by inletting the air into the vacuum chamber. Herein we investigate how the air pressure under vacuum-unit affect the thermal contact at nanometric scale. We first discuss about the materials and experimental data obtained in a vacuum-unit followed by the contribution of air pressure on the heat transfer mechanism. An analytical model is also developed to retrieve the thermal contact resistance of the present thermal probe-sample system. The behavior of the pressure dependent thermal contact resistance is also discussed according to existing models.

## 2. Materials and method

There are two bulk samples $SiO_2$ (99.99% of purity) and Ti (99.99% of purity) obtained from Neyco in the form of disk shape with diameter 10 mm and height 2 mm. The thermal conductivities of $SiO_2$ and Ti are 1.28 W/mK and 21.7 W/mK respectively. The roughness (Ra) of surface of $SiO_2$ and Ti were assessed via topography analysis. Topography of the samples were obtained by a commercial AFM of NT-MDT. The average roughness (Ra) of $SiO_2$ and Ti were 0.56 nm and 4.17 nm respectively. A homemade set-up for thermal characterization of samples under a vacuum-unit which is shown in Fig. 1(a) is used. In this unit there is a vacuum chamber in which a vacuum of the order of 1 Pa can be established. The thermal probe is at a fixed position but the sample is allowed to move with a vertical resolution of 20 nm. Only contact (at a point) measurement is possible in this set-up. To make probe-sample contact, the sample stage moves up side toward the thermal probe. The probe temperature is controlled by a Wheatstone bridge which is depicted in Fig. 1(b).

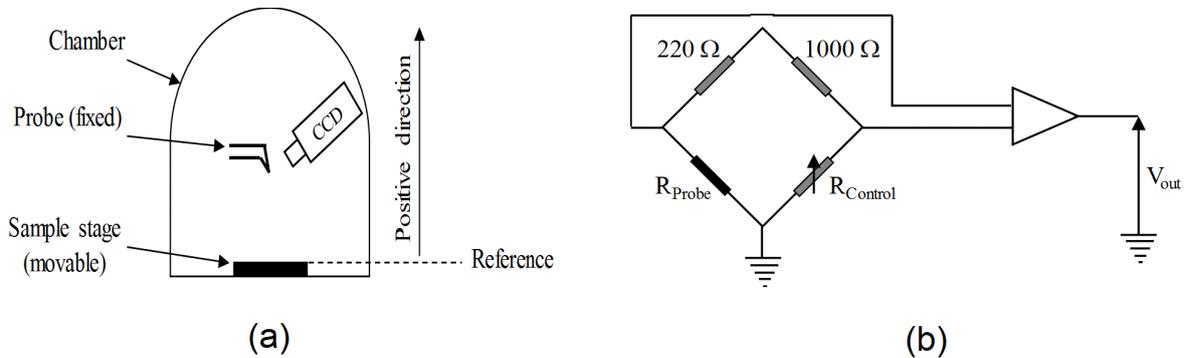

FIG. 1. (a) Schematic diagram of experimental set-up and (b) Wheatstone bridge used in the experimental set-up.

In the experiment, a Wollaston wire thermal probe from Veeco is used for the measurements of heat transfer. The Wollaston wire is made of a silver shell of 75 μm diameter and at the center there is a Platinum and Rhodium ($Pt_{90}/Rh_{10}$) alloy core of 5 μm diameter. To make thermal probe, the silver shell over a length of 200 μm is etched off electrochemically and then the bare $Pt_{90}/Rh_{10}$ wire is bent in V-shape with 15 μm radius of curvature. This $Pt_{90}/Rh_{10}$ wire is a thermo-resistive element and its temperature coefficient is 0.00165 $K^{-1}$. The thermal conductivity of the $Pt_{90}/Rh_{10}$ wire is 38 W/mK. The apex can be heated electrically by DC current[19,20]. In the present work, direct current (DC) heating mode was employed to provide the heat source to the thermal probe and the average probe temperature was fixed at 150 °C to avoid any moisture. The room temperature and humidity were maintained roughly at 20 °C and 50 % respectively throughout the experiment. We measured probe output voltage of the thermal probe at out-off-contact (oc) from sample and in-contact (ic) with sample.

### 3. Results and discussion

We recorded the output voltage of the interaction of the thermal probe to samples at four different pressures 1 Pa, 100 Pa, 5000 Pa and 100000 Pa. Fig. 2(a) and (b) display the output voltage vs probe-sample distance for $SiO_2$ and Ti respectively whereas two sub-figures on logarithmic scale are shown in Fig. 2(a) and (b) respectively. At 1 Pa, it can be observed that the output voltage is unchanged when probe moves from very far to near contact point, but there was a sudden drop in the output voltage (Fig. 2(a)) when probe comes to contact with sample. As the pressure was increased to 100 Pa (Fig. 2(b)), still there was a sudden drop in the output voltage at the point of contact and there was no significant influence of convection. But probe output voltage decreases very slowly from a far distance around 200 μm, suggesting a significant effect of air convection on the output (Fig. 2(c)). At $10^5$ Pa which is approximately atmospheric pressure there was large influence of the air medium (Fig. 2(d)).

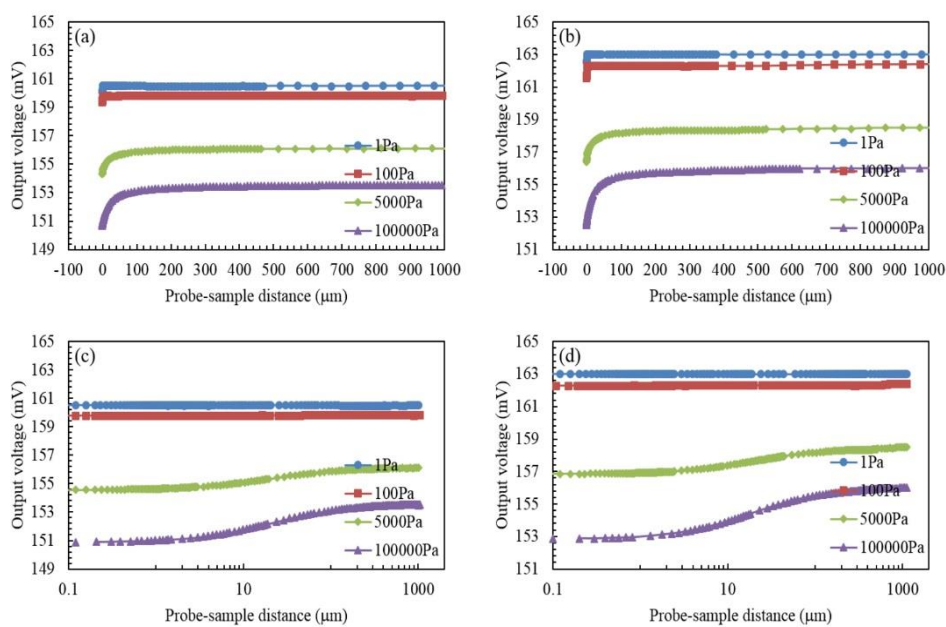

FIG. 2. Thermal probe output voltage vs probe-sample distance at different pressure 1 Pa, 100 Pa, 5000 Pa and 100000 Pa for (a). $SiO_2$ and (b) Ti. In sub-figures (c) and (d), the respective figures are shown on log x-axis.

Fig. 3(a) displays the output voltage difference ($\Delta V$ (mV) = $V_{oc} - V_{ic}$) of the thermal probe when the probe is out-off contact ($V_{oc}$) and in-contact ($V_{ic}$) whereas the normalized output voltage difference ($\Delta V/V_{oc}$) of the thermal probe is depicted in Fig. 3(b). Measurements and instruments errors are included in the error bars shown in Fig. To normalize the voltage difference, the voltage out-off contact ($V_{oc}$) is used because it is least affected output voltage. As shown in Fig. 3(a) and (b), the air pressure increases the voltage difference between out-off contact and in-contact increases and hence heat transfer increases.

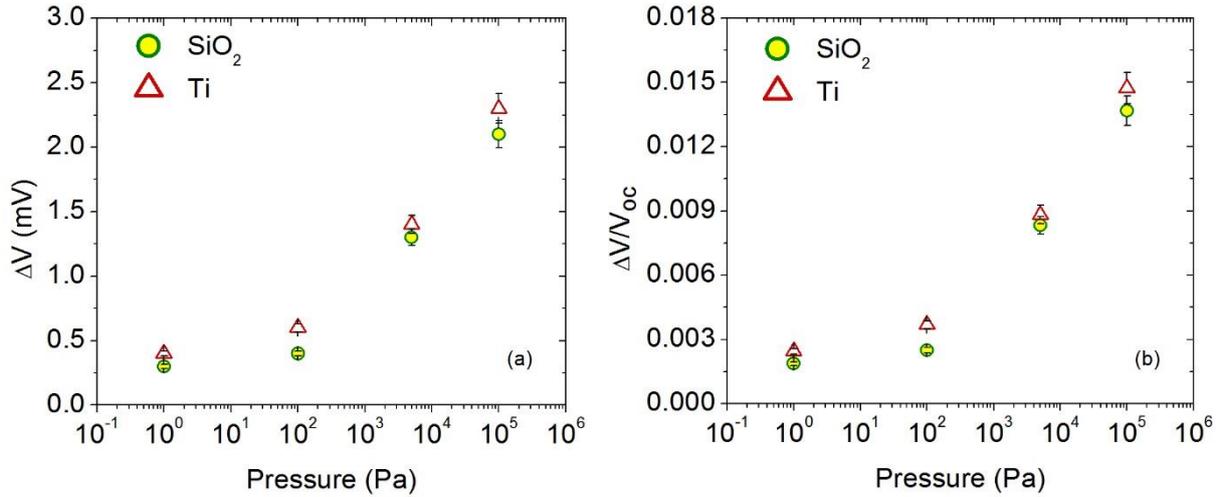

FIG. 3(a) The difference of the output voltages ($\Delta V$) and (b) difference of the normalized output voltage ($\Delta V/V_{oc}$) with respect to increasing air pressure.

To gain further insights into heat transfer mechanism at effective contact (i.e. solid-solid contact, solid-air contact and solid-water meniscus contact) between the thermal probe and sample, we have assessed thermal contact resistance within the framework of an approximate analytical model which is presented herein. Generally, at 150 °C, the water meniscus is disappeared[17,19] and also the contribution of thermal radiation on heat transfer mechanism is neglected in SThM[3,17]. However, in the previous section, the signature of normalized output voltage difference between out-off contact and in-contact was established. As a consequence, to extract the probe-sample thermal contact resistance, we propose an analytical model based on the normalized output voltage difference. In our approach to experimental setup, the same temperature ($T_0$) of the sample used as heat sink and the probe base were taken into account which is shown in Fig. 4(a). Usually, the SThM output voltage (V) is related to an excess temperature ($\Delta T$) raised by the applied power source (DC mode) and the power flow (Q[W]) through the given relation:

$$\Delta T = Q \times R = \beta \times V \qquad 1$$

Where R (K/W) is the thermal resistance of a heat transfer network and β (K/V) is a calibration factor.

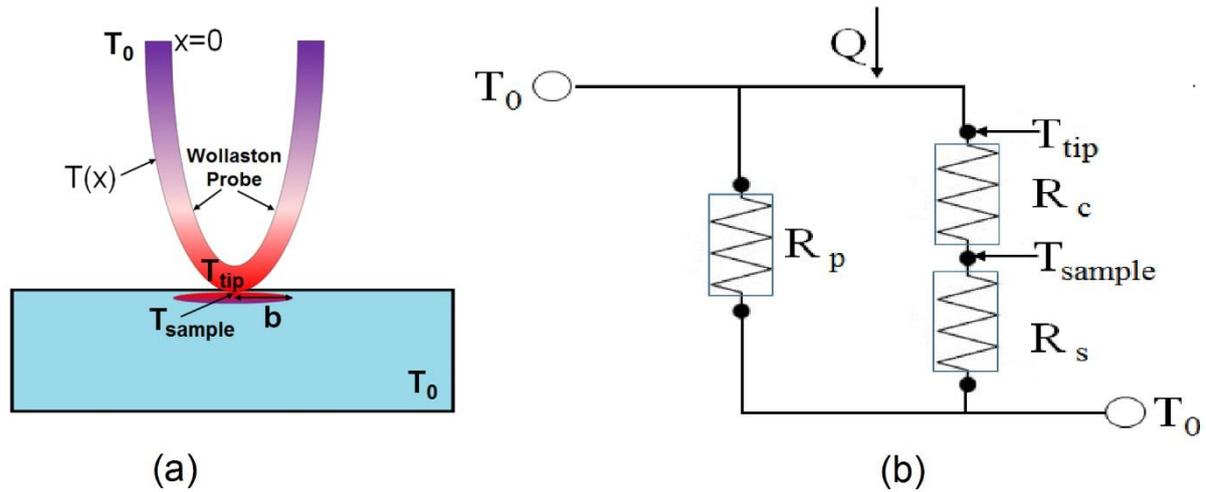

Fig. 4(a) Schematic representation of thermal interaction between thermal probe and sample (b) Schematic diagram of thermal contact resistance network when probe contacts with sample.

To develop this analytical model two hypotheses are considered. First, when the probe is out-off-contact with sample, thermal network is formed with contribution of probe itself. Thus the excess temperature ($\Delta T_{oc}$) raised in this network can be expressed as:

$$\Delta T_{oc} = Q \times R_P = \beta \times V_{oc} \qquad 2$$

where $R_P$ is the thermal resistance of probe. Second, when the probe makes contact with the sample, thermal network consists of the probe, probe-sample contact medium and sample (see Fig. 4(b)). Considering both the base of the probe and the sample heat sink at the same temperature, the excess temperature ($\Delta T_{oc}$) for in-contact thermal network can be represented as:

$$\Delta T_{ic} = Q \times R_{eq} = \beta \times V_{ic} \qquad 3$$

where $R_{eq}$ is the effective thermal resistance made of the thermal contact resistance, the sample thermal resistance. Moreover, like electrical resistances, the thermal contact resistance and the sample resistance are in series, while the thermal resistance of probe connects parallel with the thermal contact resistance and the sample thermal resistance. Therefore, $R_{eq}$ can be expressed as:

$$\frac{1}{R_{eq}} = \frac{1}{R_p} + \frac{1}{R_c + R_s} \qquad 4$$

However, by using equation 2 and 3, the normalized SThM output voltage difference can be written as below:

$$\frac{V_{oc} - V_{ic}}{V_{oc}} = \frac{\Delta T_{oc} - \Delta T_{ic}}{\Delta T_{oc}} = \frac{R_p - R_{eq}}{R_p} \qquad 5$$

Inserting the Eq. 4 into Eq. 5, the result leads to:

$$\frac{V_{oc} - V_{ic}}{V_{oc}} = \frac{R_p}{R_p + R_c + R_s} \qquad 6$$

Even though $R_c$ is considered, another important parameter thermal exchange radius should be taken into account in order to achieve the maximum accuracy of heat transfer at nano and/or micro level[17,21], because in the SThM experiments heat transfers from the probe to the sample through the circular probe-sample contact area depicted in Fig. 4(a). Thermal exchange radius can be significant in heat transfer mechanism owing to its different values with different thermal conductivity samples[21]. Considering the semi-infinite medium assumption for bulk like thickness samples, the thermal exchange radius is given by[21,22].

$$R_s = \frac{1}{4k_s b} \qquad 7$$

where $k_s$ is the thermal conductivity of the sample. For Wollaston probe, the value of b was found to be constant ($b = 2.8 \pm 0.3$ µm) for samples with thermal conductivity values ($< 2$ W/mK) but another constant value ($b = 428 \pm 24$ nm) was obtained beyond that thermal conductivity value[21]. Thus the Eq. 6 can be modified as:

$$\frac{\Delta V}{V_{oc}} = \frac{V_{oc} - V_{ic}}{V_{oc}} = \frac{R_p}{R_p + R_c + \frac{1}{4k_s b}} \qquad 8$$

According to the Eq. (8), knowing the values of $R_p$ and $k_s$, it is possible to determine the value of $R_c$. The evaluation of $R_c$ as a function of pressure is displayed in the Fig. 5. The estimated $R_c$ shows the range between ~$1.8 \times 10^7$ K/W and ~$14.3 \times 10^7$ K/W for $SiO_2$ and Ti materials. B. Gotsmann *et al.*[9] reported $R_c$ value of ~$10^7$ K/W under vacuum whereas many others works insure $R_c$ values with the range between ~$10^4$ K/W and ~$10^9$ K/W for different systems[17,21,23–25]. The study suggests that our estimated of $R_c$ values lie in the range of values reported in the literatures. Therefore, the obtained $R_c$ values strongly validates the analytical model which is presented here.

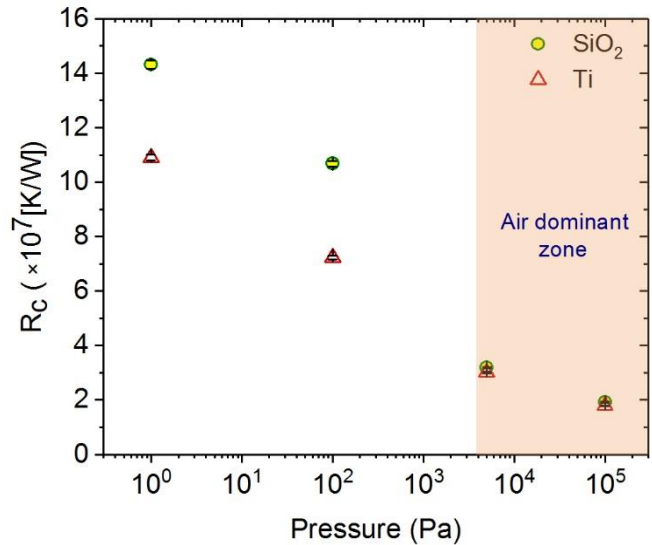

FIG. 5. Thermal contact resistance for $SiO_2$ and Ti as a function of Pressure at probe temperature 150°C.

Furthermore, the result reveals that the $R_c$ systematically increases with decreased air pressure. In 2015, G. T. Hohensee *et al.*[18] reported that metal-diamond interface thermal resistance increases with decreased mechanical pressure (~0-50 GPa). Wollaston probe-sample contact resistance also exhibit similar behaviour in air pressure. Based on effective contact between probe and sample, such behaviour of the thermal contact resistance can be interpreted. We suggested that heat transport is not only dependent on solid-solid (probe-sample) contact, but also it is intimately tied to the air confined within asperities which act as heat transfer medium. Fig. 6 presents a schematic view of the heat transfer from the probe to the sample through the probe-sample contact area. At high pressure, the heat transfer through air contributes significantly on the thermal contact resistance (see Fig. 6(a) and (b)). Also the signature of thermal contact resistance insures that heat transfer through effective contact medium is reduced due to induced void medium which is created by elimination of the air pockets under low pressure (see Fig. 6(c) and (d)). In a sense, heat transfer is somewhat impeded at that pressure, because air molecules (heat carriers) become rare. Therefore, the overall thermal contact resistance increases under low pressure.

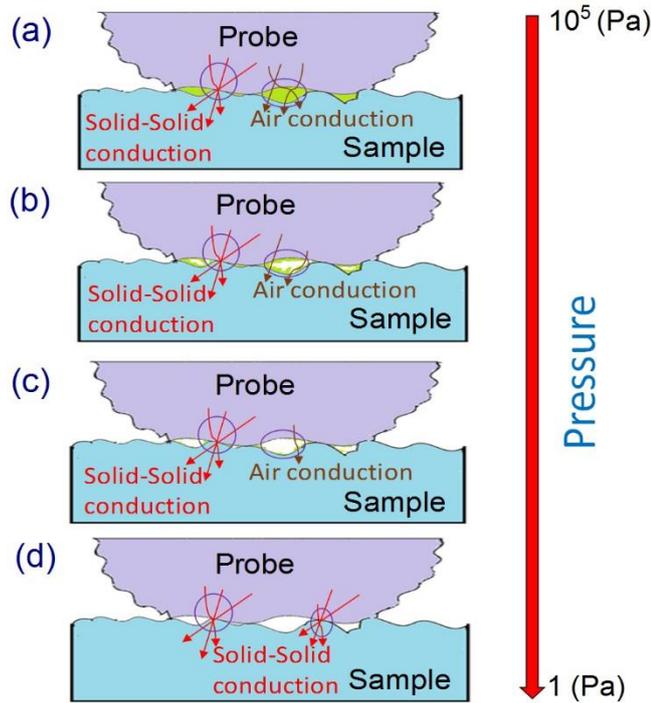

FIG. 6. Schematic presentation of thermal interaction between probe and sample.

Alternatively, the obtained $R_c$ signature can be explained by using the concept of thermal circuit like electrical analogue. The effective contact resistance can be presented as $\frac{1}{R_c} = \frac{1}{R_{air}} + \frac{1}{R_{solid-solid}}$, where $R_{air}$ and $R_{solid-solid}$ are thermal resistance of air and probe-sample contact respectively. Although pressure is changed herein, the probe-sample contact space is unchanged. Therefore, $R_{solid-solid}$ is considered as constant throughout the pressure. Moreover, for a rough estimation of thermal conductivity of air as a function of pressure, the given equation is used [26]:

$$k_{air} = k_0 \times \frac{1}{1+\frac{7.6\times10^{-5}}{P\times\frac{D}{T}}} \quad (W/mK) \qquad 9$$

where $k_0$, T, D and P represent the thermal conductivity of air at room pressure and temperature, the average temperature of probe-sample system, the length of cavity induced at Wollaston probe-sample contact area when the probe is in-contact mode and the pressure. Since heat transfers through nano-contact area, we assume that a void space (cavity) of 100 nm is induced between probe and sample where air is confined. Also having T=423K and $k_o$=0.0284 WmK, the thermal conductivity of air is approximated as function of pressure which is displayed in Fig. 7. The figure exhibits a decreasing behaviour of the thermal conductivity of air with decreased pressure, which is consistent with those reported in literatures [26,27]. It means that the resistance increases due to weak dependence of air at low pressure. Thus the effective thermal contact resistance ($R_c$) increases with decreased pressure. So the explanation supports the above behaviour of pressure dependent thermal contact act resistance.

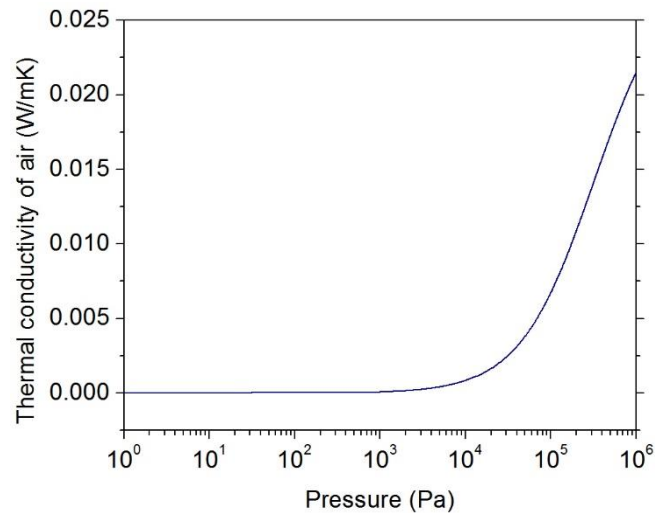

FIG. 7. Estimated thermal conductivity signature of air as a function of pressure at 150 °C.

In addition, as shown in Fig. 5, the thermal contact resistance of $SiO_2$ and Ti are almost similar at high pressure which is found to be consistent with previous reported literature[21]. The result indicates that air strongly dominates the effective thermal contact resistance under high pressure (at 5000 Pa and $10^5$ Pa). But under low air pressure condition, there is difference between the thermal contact resistance of $SiO_2$ and Ti. Since air pockets are eliminated at low pressure, the heat transfer takes place only through solid/solid contact which supports the above explanation. This difference of the thermal contact resistance can be explained by thermal mismatch in thermal conductivity of probe and sample. The difference of the thermal conductivity ($\Delta k = k_p - k_s$; where $k_p$ is the thermal conductivity of the probe and $k_s$ is the thermal conductivity of the sample) between probe and $SiO_2$ is greater than the difference of the thermal conductivity between probe and Ti. Thus large thermal mismatch between probe and $SiO_2$ results in large thermal contact resistance in compare with probe and Ti where thermal mismatch is relatively less. To address this issue, we include a calculation of transmission coefficient of the phonon ($\alpha$) and thermal contact conductance ($G[W/m^2K]$) between two dissimilar materials using acoustic mismatch model (AMM) and diffuse mismatch model (DMM). which are highlighted bellow.

Using the Landauer formalism, per unit area thermal contact conductance (G) between two materials A (heat source) and B is given by the following expression[28]:

$$G = \frac{1}{4}\sum_j \int_0^{\omega_{A,j}^v} D_{A,j}(\omega)\frac{\partial n(\omega,T)}{\partial T}\hbar\omega v_{A,j}\alpha_{A\to B,j}(\omega)d\omega, \qquad 10$$

where $\omega$ is phonon frequency, $D$ is phonon density of states, $n$ is the Bose-Einstein distribution function and $v$ is phonon velocity. The subscript "$j$" is for polarization of the phonon. In the AMM and DMM approach, the phonon transmission coefficients are given by[29,30]:

$$\alpha_{AMM,A\to B} = \frac{\rho_A v_A \rho_B v_B}{(\rho_A v_A + \rho_B v_B)^2} \qquad 11$$

$$\alpha_{DMM,A\to B} = \frac{\sum_j D_{B,j} v_{B,j}}{\sum_j D_{A,j} v_{A,j} + \sum_j D_{B,j} v_{B,j}} \qquad 12$$

To estimate the approximate thermal contact conductance at the interface of the Wollaston wire thermal probe and sample, we consider a platinum wire instead of $Pt_{90}/Rd_{10}$ wire. We assume that materials Pt, Ti and $SiO_2$ are homogeneous and isotropic. The data of Pt, Ti and $SiO_2$ are taken from the literature[31–33] and the values of all required parameter used in the models are listed in Table I.

TABLE I: Parameters used in the AMM and DMM model to calculate thermal contact resistance between Wollaston wire and sample. Materials have been considered as isotropic.

| Material | Average phonon velocity, $v$ (m/s) | Density, $\rho$ (kg/m$^3$) | Number density, N |
|---|---|---|---|
| Pt [31] | 2599 | $21.09 \times 10^3$ | $6.51 \times 10^{28}$ |
| Ti [32] | 4012 | $4.51 \times 10^3$ | $5.67 \times 10^{28}$ |
| SiO$_2$ [33] | 4169 | $2.65 \times 10^3$ | $2.66 \times 10^{28}$ |

Using the average sound velocities in the materials, the values of $\alpha$ and G are determined which is listed in Table II. As shown in Table II, the AMM model gives slightly higher value than the DMM model as it considers the elastic interaction of the phonon. The calculation shows that t and G values are higher for Ti than the SiO$_2$. Also in Table 1, the thermal contact resistance per unit area (R = 1/G) is shown. These values have same order (10$^{-9}$) as reported in the literature[30]. The study indicates the heat transport through Wollaston probe-Ti interface is better than the probe-SiO$_2$ interface. Thus, the result evidences the above explanation. We therefore believe that our estimation of $R_c$ can provide more light in nanoscale heat transfer mechanism for better understanding and also may open up an additional pathway to improve heat dissipation in probe-sample system.

TABLE II: Transmission coefficient of the phonon ($\alpha$), per unit thermal contact conductance ($G$) and per unit thermal contact resistance ($R$) as obtained by the AMM and DMM model at 150 °C temperature.

| Sample | AMM | | | DMM | | |
|---|---|---|---|---|---|---|
| | $\alpha$ | G(W/m$^2$K) | R(m$^2$K/W) | $\alpha$ | G(W/m$^2$K) | R(m$^2$K/W) |
| Ti | 0.75 | $8.49 \times 10^8$ | $1.18 \times 10^{-9}$ | 0.30 | $3.36 \times 10^8$ | $2.97 \times 10^{-9}$ |
| SiO$_2$ | 0.56 | $6.35 \times 10^8$ | $1.58 \times 10^{-9}$ | 0.28 | $3.18 \times 10^8$ | $3.14 \times 10^{-9}$ |

## 4. Conclusion

In conclusion, the thermal interaction of the thermal probe with $SiO_2$ and Ti bulk samples has been investigated under vacuum. SThM was employed to measure the difference of output voltages of the thermal probe at out-off-contact and in-contact position. The result reveals that the difference of output voltages decreases with increasing air pressure which suggests an enhancement of the heat transfer at nanoscale. In addition, a simple analytical model has been developed to estimate the thermal contact resistance between the thermal probe and samples by using the normalised output voltage difference. The result exhibits that $R_c$ decreases with increasing air pressure. This signature of $R_c$ is interpreted in the context of heat transfer through air confined within probe-sample nano-contact. As the amount of air increases it fills the voids at the interface of the thermal probe and sample. Denser air has greater thermal conductivity and hence lesser thermal contact resistance between the thermal probe and bulk samples. The explanation is also supported in the context of the thermal mismatch behaviour which is studied by using acoustic mismatch model and diffuse mismatch model.

## 5. Acknowledgements

Authors acknowledges the European Union for financial support through European Union Seventh Framework Programme FP7-NMP-2013-LARGE-7 under grant agreement no.604668 (QuantiHeat project). The authors would like to thank Didier Caron for helpful technical support.

**Figure Captions**

FIG. 1. (a) Schematic diagram of experimental set-up and (b) Wheatstone bridge used in the experimental set-up.

FIG. 2. Thermal probe output voltage vs probe-sample distance at different pressure 1 Pa, 100 Pa, 5000 Pa and 100000 Pa for (a). $SiO_2$ and (b) Ti. In sub-figures (c) and (d), the respective figures are shown on log x-axis.

FIG. 3(a) The difference of the output voltages ($\Delta V$) and (b) difference of the normalized output voltage ($\Delta V/V_{oc}$) with respect to increasing air pressure.

Fig. 4(a) Schematic representation of thermal interaction between thermal probe and sample (b) Schematic diagram of thermal contact resistance network when probe contacts with sample.

FIG. 5. Thermal contact resistance for $SiO_2$ and Ti as a function of Pressure at probe temperature 150°C.

FIG. 6. Schematic presentation of thermal interaction between probe and sample.

FIG. 7. Estimated thermal conductivity signature of air as a function of pressure at 150 °C.

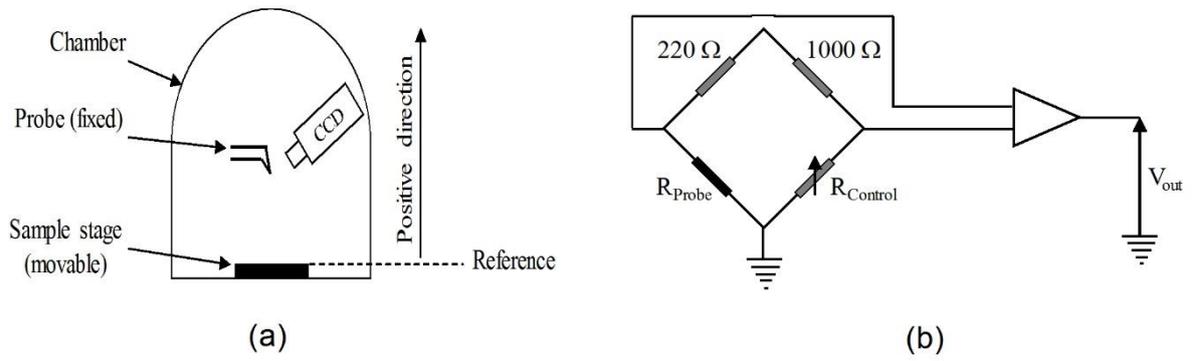

FIG 1

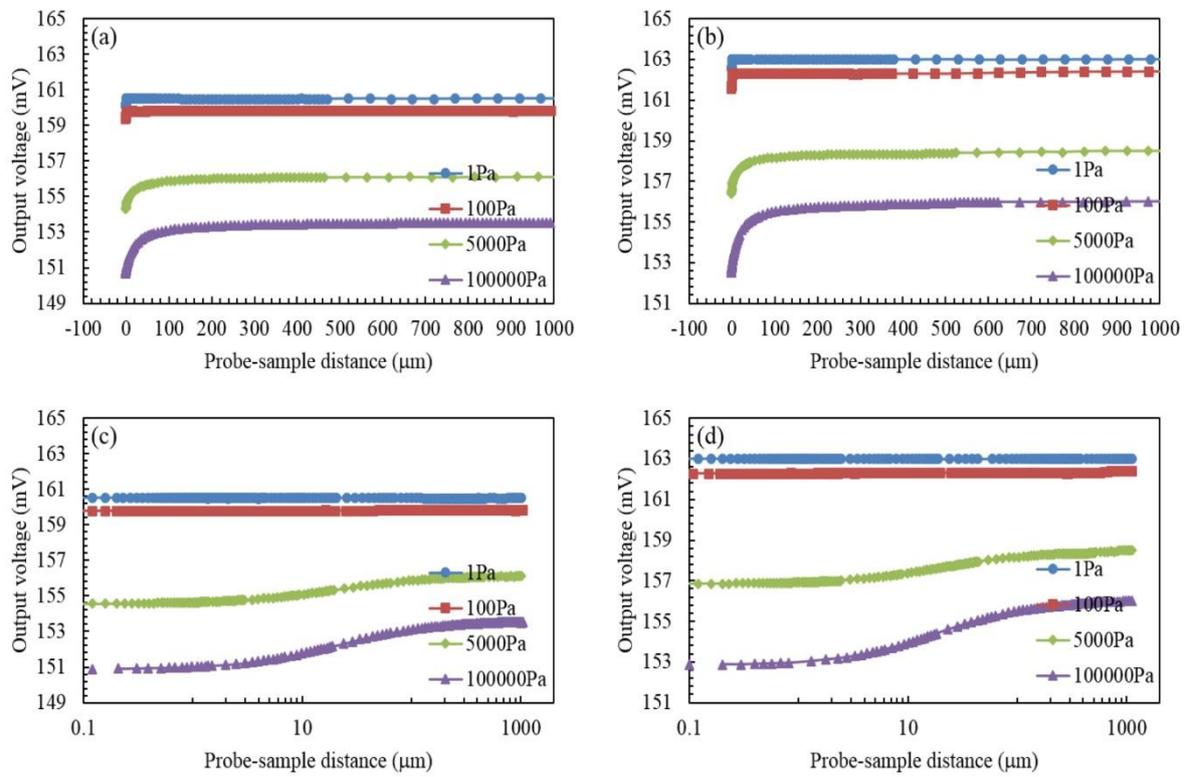

FIG 2

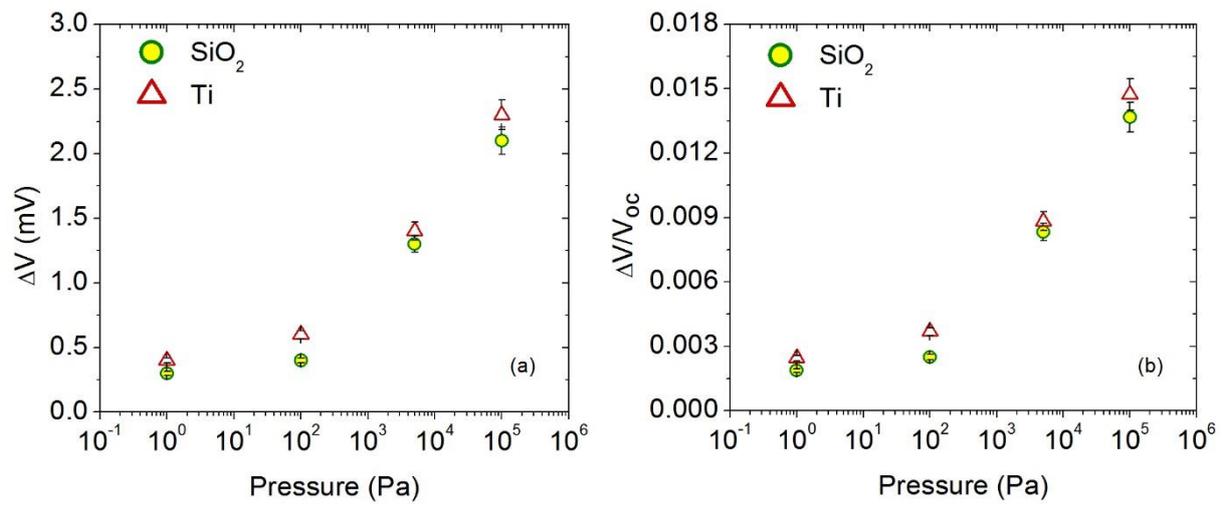

FIG 3

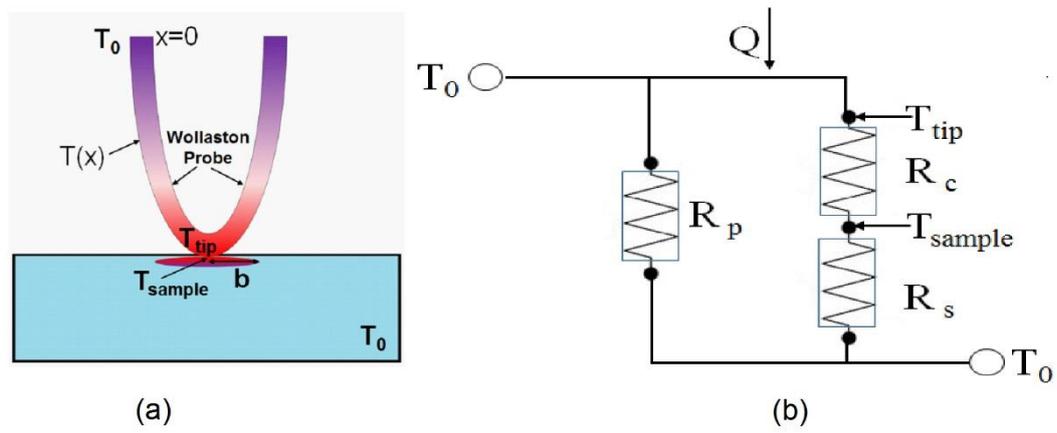

FIG 4

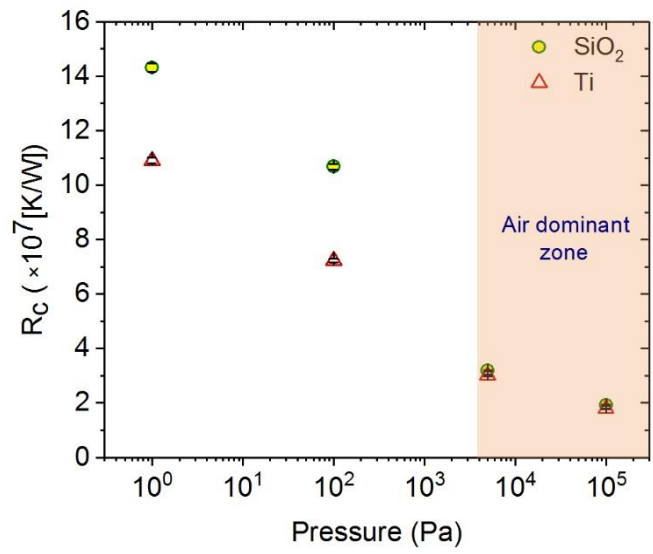

FIG 5

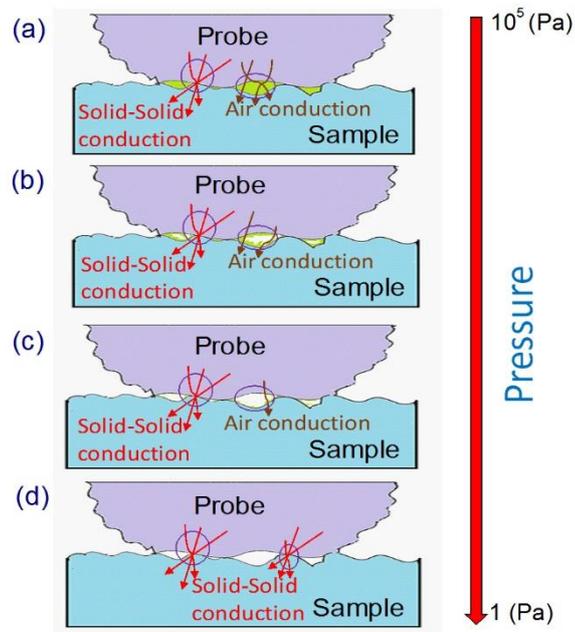

FIG 5

FIG 6

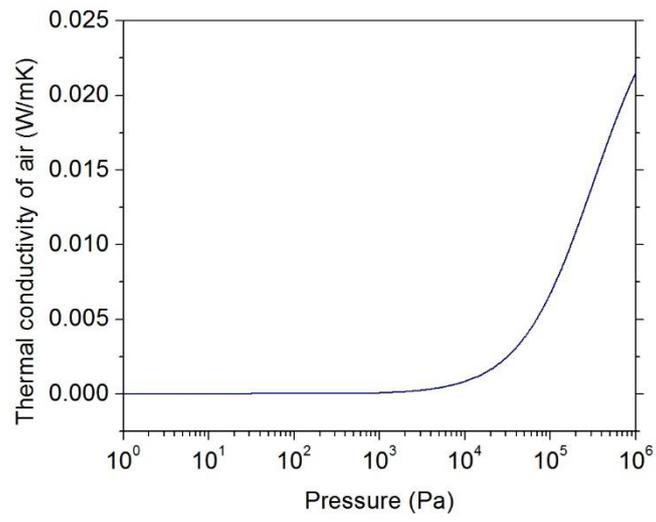

FIG 7